\documentclass[prl, reprint, nofootinbib, showpacs,preprintnumbers,amsmath,amssymb]{revtex4}
\usepackage{varioref,exscale,latexsym,amsmath,amssymb}
\usepackage{graphicx,tikz}

\usepackage{slashed}
\usepackage{dcolumn}
\usepackage{bm}
\usepackage{hyperref}
\usepackage{color}
\usepackage{xcolor}
\usepackage{booktabs}
\usepackage[normalem]{ulem}
\usepackage{braket}

\newcommand{\beq}{\begin{equation}}
\newcommand{\eeq}{\end{equation}}
\newcommand{\bea}{\begin{eqnarray}}
\newcommand{\eea}{\end{eqnarray}}

\def\lsi{\raise0.3ex\hbox{$<$\kern-0.75em\raise-1.1ex\hbox{$\sim$}}}
\def\gsi{\raise0.3ex\hbox{$>$\kern-0.75em\raise-1.1ex\hbox{$\sim$}}}

\def\cM{\mathcal{M}}

\newcommand{\be}{\begin{equation}}
\newcommand{\ee}{\end{equation}}

\def\cL{{\cal L}}
\def\cM{{\cal M}}

\DeclareMathOperator\arccot{arccot}

\begin{document}
\preprint{ACFI-T23-xx}

\title{\bf 
Amplitudes and Renormalization Group Techniques: A Case Study }

\medskip\

\medskip\

\author{Diego Buccio${}^1$, John F. Donoghue${}^2$ , Roberto Percacci${}^1$}
\email{donoghue@physics.umass.edu}
\affiliation{
${}^1$International School for Advanced Studies,
Via Bonomea 265, 34134 Trieste, Italy\\
and INFN, Sezione di Trieste, Trieste, Italy\\
${}^2$Department of Physics,
University of Massachusetts,
Amherst, MA  01003, USA
}

\begin{abstract}
We explore the properties of a simple renormalizable shift symmetric model 
with a higher derivative kinetic energy and quartic derivative coupling,
that can serve as a toy model for higher derivative theories of gravity.
The scattering amplitude behaves as in a normal effective field theory
below the threshold for the production of ghosts,
but has an unexpectedly soft behavior above the threshold.
The physical running of the parameters is extracted from 
the 2-point and 4-point amplitudes. 
The results are compared to those obtained by other methods
and are found to agree only in limiting cases.
We draw several lessons that may apply also to gravity.
\end{abstract}

\maketitle

\section{1. Introduction}

There are various ways of understanding the notion of ``running coupling''
and the associated Renormalization Group (RG),
and they do not always give equivalent results.
We can distinguish at least three different notions.
The one that is of greatest significance in particle physics arises when one identifies the running coupling by a measurement of a physical amplitude, at a renormalization scale $\mu_R$. 
We will refer to this as “physical running”.
It is in this sense that one understands, for example, the running of the couplings of the standard model.
The second notion is the dependence of the renormalization of the coupling on a scale $\mu$ 
that is introduced to fix dimensions, such as factors $\mu^{4-d}$ in dimensional regularization, 
or $\log(\Lambda/\mu)$ if cutoff regularization is used.
This is a less physical notion, because such a  scale is just an intermediate device that does not appear in physical observables.
In particular, $\mu$ is conceptually different from the renormalization scale $\mu_R$.
However, it is often useful because it can be easier to calculate,
and in the right circumstances (that we shall discuss in more detail later)
it is a good proxy for the momentum dependence.
Finally, there is a more general notion of ``running'',
defined as the dependence of a coupling on some external mass or energy scale,
typically an UV or IR cutoff, where one often considers the simultaneous running of many - even infinitely many - couplings, collected into some kind of potential or action.
This originated with Wilson's RG \cite{Wilson:1973jj},
and its generalizations \cite{Wegner:1972ih,Polchinski:1983gv}.
A more recent version of it that applies to a scale-dependent 1PI effective action
will be referred here as the Functional RG (FRG) \cite{wett1,Morris:1993qb}.
These are much more broader definitions of RG, but in some particular cases
can be related to the other two.
They have proven very useful in the understanding of critical phenomena
and in other contexts, see \cite{Dupuis:2020fhh} for a recent review.

One of the goals of this paper is to illustrate the relations between these notions
in a very simple model consisting of a single scalar with Lagrangian:
\footnote{We are using the metric signature $(-,+,+,+)$.}
\be
\cL=
-\frac12 Z_1(\partial\phi)^2
-\frac12 Z_2(\Box\phi)^2
-\frac14 g((\partial\phi)^2)^2\ .
\label{action1}
\ee
It has a shift symmetry $\phi\to \phi + {\rm constant}$ 
and reflection symmetry $\phi\to -\phi$,
and is renormalizable despite the derivative interaction term. 
The higher derivative interactions are similar to those which appear in gravitational theories. The inclusion of both two and four derivatives in the kinetic energies is typical of many applications of the FRG in which operators of different dimensions appear. It is also a crucial part of Lee-Wick theories \cite{Lee:1970iw} and of Quadratic Gravity \cite{Salvio:2018crh, Donoghue:2021cza}.

Superficially, this theory appears to be pathological.
First, theories with higher derivative kinetic energies contain extra degrees of freedom. 
This can be seen from the propagator in the full theory 
(with $Z_1=1$ and $Z_2=1/m^2$)
\beq
iD_F(q^2) = \frac{-i}{p^2+\frac{p^4}{m^2} } = -i\left[ \frac1{p^2} - \frac1{p^2 +m^2}\right]\ .
\label{parfrac}
\eeq
We see that besides a normal massless particle,
there is also a ghost with mass $m$.
The signs in the original Lagrangian have been chosen to have the massive pole at timelike momentum, because the opposite sign would have the pole being tachyonic.
Second, as we shall discuss, the theory is asymptotically free, {in the sense that the coupling runs logarithmically to zero in the UV limit}, but only for negative coupling.
In this it is reminiscent of Symanzik's observation that ordinary $\lambda\Phi^4$
theory is asymptotically free for $\lambda<0$ \cite{Symanzik:1973hx}.
In spite of this, we can study the renormalization of this model
and draw from it some useful lessons.

Besides being an interesting case study for several aspects of renormalization theory,
our model is also of independent interest and has appeared recently in various different contexts.
Without the higher derivative kinetic term,
it is a textbook example of Effective Field Theory (EFT),
being the low energy description of a $U(1)$-invariant linear sigma model
in the Higgs phase \cite{Burgess:2020tbq,Donoghue:2022azh}.
With the higher derivative kinetic term, it is the low energy EFT
for the higher derivative version of the same model.
As a CFT, the higher-derivative model has been discussed in \cite{Safari:2021ocb}.
In the context of Asymptotic Safety, it has been presented as a type
of matter interaction that would necessarily have to be present
if gravity has a nontrivial fixed point \cite{deBrito:2021pyi,Laporte:2021kyp}.
It has also been treated by the full machinery of the FRG by two of the present authors
\cite{Buccio:2022egr}, viewing it as a toy model for gravity.
It was found to run from the higher derivative free fixed point at high energy
to the standard free fixed point at low energy.
Finally, it has been studied recently by Tseytlin \cite{Tseytlin:2022flu}
and by Holdom \cite{Holdom:2023usn},
who found evidence that the model may be less pathological than
would first appear.

We will always assume that the field {$\phi$} has mass dimension one,
which is the natural choice when we interpret the two-derivative term
as defining the propagator.
Then, $Z_2$ and $g$ have dimension of inverse mass to power 2 and 4 respectively.
It is thus natural to reparametrize
\beq
\label{notation}
\cL =-\frac{Z_1}{2} \partial_\mu \phi \partial^\mu\phi - \frac{Z_1}{2m^2}\Box\phi \Box \phi - \frac{Z_1^2g}{4M^4}( \partial_\mu \phi \partial^\mu\phi )( \partial_\nu  \phi \partial^\nu \phi)
\eeq
where $m$ and $M$ are masses and $g$ is dimensionless.
Moreover, we have defined the coupling constant with an explicit factor of the mass $M$ in order to make $g$ dimensionless. The value of this somewhat redundant notation is that it facilitates the use of dimensional analysis by showing the mass factors explicitly. 
The notation is natural when one views this as the low energy
limit of the $U(1)$ linear sigma model,
in which case the masses $m$ and $M$ are parametrically independent.
Even though here we shall consider the theory as being potentially UV complete
in itself, without the radial mode, we shall retain this notation.
One can set $M=m$ without loss of generality.

In order to see this as a toy model for gravity,
we recall that the action of Quadratic Gravity is schematically of form
$m_P^2 R+\tfrac{1}{\xi}C^2$ (with $C$ the Weyl tensor), 
so if we rescale the metric fluctuation by $m_P\sim 1/\sqrt{G}$,
the action contains, among other terms,
$$
(\partial h)^2
+\frac{1}{\xi m_P^2}(\Box h)^2
+\frac{1}{\xi m_P^4}(\partial h)^4\ .
$$
Recalling that the mass of the ghost is $m=\xi m_P^2$,
this becomes essentially the same as (\ref{notation})
with $Z_1=1$, $M=m$ and $g=\xi$.

Irrespective of the notation, it is important to keep in mind that
the Lagrangian contains two mass scales:
the mass of the ghost, $m$, and the scale $M/\sqrt[4]{g}$
at which tree level unitarity is violated
and above which one would appear to be in a strongly interacting regime, {due to the $E^4$ derivative interaction}.
In this paper we will always assume that $m<M/\sqrt[4]{g}$,
in such a way that the massive ghosts can propagate and
still be weakly interacting.
Depending on the characteristic scale of the process,
we thus have three energy regions which have different behavior. 
Let us name these:
\begin{itemize}
\item {\bf Low Energy (LE):} This region is defined by energies small compared to the ghost mass $m$. The heavy ghost is not dynamically active and can be integrated out.
\item  {\bf Intermediate Energy (IE):} This corresponds to energies above the mass $m$, but below the apparent strongly-interacting regime. Here the heavy ghost is dynamically active. In an appendix we will also briefly comment on an intermediate case where $s\sim -u>>m^2$ but $t<<m^2$.
\item {\bf High Energy (HE):} This occurs when the energy is high enough that $gE^4/M^4 > 1$. At these energies, perturbation theory would seem to break down.
\end{itemize}

In this paper we will study the scattering amplitude of this theory
in the first two regimes.
At tree level it is given by
\be
-\frac12 g(s^2+t^2+u^2)\ .
\label{treeamp}
\ee
At low energy, quantum corrections generate new effective interactions
with six or eight derivatives.
This is the expected behavior of a nonrenormalizable theory,
treated with standard EFT methods.
Somewhat unexpectedly, the higher dimension operators
cancel above the mass threshold for the production of ghosts,
leaving us with a theory that looks renormalizable,
with a logarithmically running coupling.

By comparing the loop corrections with the two- and four-point amplitudes, 
we will identify the physical running (or lack of running) of the parameters. 
The results differ in general from those given by 
other methods, but agree in some limits.

We will proceed as follows.
We begin by describing the model in the absence of the higher derivative kinetic term.
It is useful to have this description because the full theory reduces to this EFT 
in the low energy limit. 
In Sections 3 and 4 the calculation of the two-point function and the four-point scattering amplitude are presented.  The final results are given in Eq. (\ref{2point}) for the two point quantum correction, and in Eq (\ref{4point}) for the full four point correction.
Section 5 is devoted to an analysis of the amplitude.
In particular we will see how to match it at low energy to the previously obtained EFT
results, and also consider the remarkable simplifications that occur
in the high energy limit, Eqs. (\ref{smallS}) and (\ref{bigS}).  
In section 6 we compare the physical beta functions derived from the amplitude
to the beta functions obtained from the FRG and other definitions. 
Our main conclusions are summarized in Section 7.

\section{2. Effective Field Theory at Low Energy.}

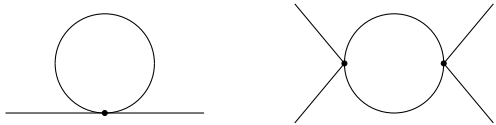
\begin{figure}[ht]
\begin{center}
\begin{tikzpicture}
[>=stealth,scale=0.66,baseline=-0.1cm]
\draw (-1,0) arc (180:0:1cm);
\draw (-1,0) arc (180:360:1cm);
\draw (2,-1) node[anchor=west] {} -- (0,-1);
\draw (0,-1) -- (-2,-1) node[anchor=east] {};
\draw[fill] (0,-1) circle [radius=0.05];
\end{tikzpicture}
\qquad
\begin{tikzpicture}
[>=stealth,scale=0.66,baseline=-0.1cm]
\draw (-1,0) arc (180:0:1cm);
\draw (-1,0) arc (180:360:1cm);
\draw (2,1.2) node[anchor=west] {} -- (1,0);
\draw (2,-1.2) node[anchor=west] {} -- (1,0);
\draw (-1,0) -- (-2,1.2) node[anchor=east] {};
\draw (-1,0) -- (-2,-1.2) node[anchor=east] {};
\draw[fill] (-1,0) circle [radius=0.05];
\draw[fill] (1,0) circle [radius=0.05];
\end{tikzpicture}
\caption{The diagrams 
giving corrections to the two- and four-point functions.}
\label{fig.chiral}
\end{center}
\end{figure}

In generating an Effective Field Theory (EFT) one needs to know the low energy degrees of freedom and the symmetries. The massless mode is the only one which is dynamical at low energy. The symmetry is the same as that of the full theory, which in this case consists of the shift and reflection symmetry.  One then writes out a normal theory with only the massless particle, consistent with these symmetries. In general this may have higher derivative nonrenormalizable interactions. By this procedure we arrive at the Lagrangian
\beq
{\cal L} = -\frac{1}{2} \partial_\mu \phi \partial^\mu\phi - \frac{g}{M^4}( \partial_\mu \phi \partial^\mu\phi )( \partial_\nu  \phi \partial^\nu \phi)  + {\cal L}_6 +{\cal L}_8 + ... \ \ .
\eeq
Here  ${\cal L}_6$ and ${\cal L}_8$ are Lagrangians with six and eight derivatives, which will be described more fully below.  In principle one might consider a notation where the coupling strength $g$ differs from that of the original theory. However we will see that the coupling in the effective theory is identified with the coupling of the full theory when the latter is renormalized at low energy. 

At one loop, wavefunction renormalization would arise from the tadpole diagram with two external legs, as shown in the two-point diagram of Figure \ref{fig.chiral} However this vanishes, because the tadpole integral
\beq
\int \frac{ d^d k}{(2\pi)^d} \frac{ k_\mu k_\nu}{k^2} 
\eeq
is a scale-less integral which vanishes in dimensional regularization. This sets $Z_1=1$ in the effective field theory limit. 

For the scattering amplitude, the one loop amplitude arises at order $E^8$ or equivalently it is described by a Lagrangian with eight derivatives. This can be seen dimensionally from the factor of $g^2/M^8$ which arises from two factors of the fundamental interaction. In dimensional regularization there are no other mass scales in the theory, and so the numerator factors arising from the one loop amplitude must be powers of the external energies. There will be a divergence in this amplitude and the coefficients at order $E^8$ will need to be renormalized. Along with the renormalization will come the usual logs, and because this is a mass independent renormalization, these must be factors of $\log s/\mu^2$ or similar logarithms. This tells us that the coefficients at order $E^8$ can be interpreted as ``physically running'' couplings. 
These logarithms will be finite and are predictions of the effective field theory. 

In contrast, there will be no renormalization of the coefficients at order $E^4$ or $E^6$, as seen by the power counting described in the previous paragraph. This implies that there also will not be any logarithms generated. The couplings at order $E^4$ and $E^6$ will not be running 
in the physical sense. 

Let us see this in explicit detail. In $\phi + \phi \to \phi +\phi$ scattering, there are a limited number of kinematic invariants involved consistent with the crossing symmetry of the amplitude. This limits the number of effective Lagrangians involved. At dimension six and eight, these can be taken to be
\bea
{\cal L}_6 &=&\frac{g_6}{4M^6} \partial_\mu \phi \partial^\mu \phi \Box \partial_\nu \phi \partial^\nu \phi + \frac{{g}_6'}{4M^6} \partial_\mu \phi \partial_\nu \phi \Box \partial^\mu \phi \partial^\nu \phi  \nonumber \\
{\cal L}_8 &=&- \frac{g_8}{4M^8} \partial_\mu \phi \partial^\mu \phi \Box^2 \partial_\nu \phi \partial^\nu \phi - \frac{{g}_8'}{4M^8} \partial_\mu \phi \partial_\nu \phi \Box^2 \partial^\mu \phi \partial^\nu \phi  \eea
The coupling constants in these Lagrangians cannot be predicted from effective field theory alone.

We have calculated the one loop scattering amplitude in this theory.  
From the explicit calculation, the $s$ channel gives
\be
\frac{ig^2 s^2 \left(41 s^2+t^2+u^2\right)}{1920 \pi ^2 M^8\epsilon
   }-\frac{i g^2s^2 \left(15(\log (\frac{-s}{4\pi\mu^2}) 
+\gamma_E)\left(41 s^2+t^2+u^2\right)-1301 s^2-46 t^2-46
   u^2\right)}{28800 \pi ^2M^8}+O\left(\epsilon ^1\right),
\ee
while channels $t$ and $u$ can be found thanks to crossing symmetry. Channel $t$ is given by the substitution $s\to t$ and $t\to s$ and $u$ corresponds to the cyclic permutation $s\to u$, $t\to s$, $u\to t$.
the total one-loop quantum correction to the four-point amplitude is
\bea
&&\frac{g^2\left(41(s^4+t^4+u^4)+2(s^2 t^2+t^2 u^2+u^2s^2)\right)}{1920 \pi ^2 M^8\epsilon }
\nonumber\\
&&-\frac{g^2}{28800 \pi^2M^8}
   \bigg\{15 \bigg[s^2 \left(41 s^2+t^2+u^2\right)\log \left(\frac{-s}{4 \pi  \mu^2
   }\right) 
\nonumber \\
   &&\qquad\qquad\qquad\qquad
+t^2
   \left(s^2+41 t^2+u^2\right) \log \left(\frac{-t}{4 \pi 
   \mu^2 }\right)
\nonumber \\
   &&\qquad\qquad\qquad\qquad
+u^2 \left(s^2+t^2+41 u^2\right) \log
   \left(\frac{-u}{4 \pi  \mu^2 }\right)\bigg]
\nonumber \\
   &&-(1301-615 \gamma_E )(s^4+t^4+u^4)-2(46-15 \gamma ) (s^2 t^2+t^2 u^2+u^2s^2)\bigg\}+O\left(\epsilon ^1\right)
\eea
Because the field here is massless, the logarithms can only involve kinematic factors of $s,~t,~u$.

The divergence in this expression can be absorbed into the renormalization of the dimension 8 coefficients in the effective Lagrangian. 
When renormalized at a scale $s=t=u=\mu_R^2$,  the amplitude has the form
\bea\label{EFTamp}
{\cal M} &=&-\frac{g}{2M^4} (s^2+t^2+u^2)  \nonumber \\
&& +\frac{g_6}{2M^6} (s^3+t^3+u^3) +\frac{g_6'}{4M^6} (s^2t +s^2 u +t^2 u +t^2 s + u^2 s +u^2 t) \nonumber \\
&& -\frac{g_8(\mu_R)}{M^8} (s^4+t^4+u^4) 
-\frac{g_8'(\mu_R)}{2M^8} (s^2t^2 +s^2 u^2+t^2 u^2)  
\nonumber \\
&& -\frac{ g^2}{1920\pi^2M^8} \left[ 41 s^4 \log \left(\frac{-s}{\mu_R^2}\right)  
+41 t^4 \log\left(\frac{-t}{\mu_R^2}\right) 
+ 41 u^4 \log\left(\frac{-u}{\mu_R^2}\right)  
\right. \nonumber \\
&& +\left. s^2(t^2+u^2) \log\left(\frac{-s}{\mu_R^2}\right) 
+t^2(s^2+u^2) \log\left(\frac{-t}{\mu_R^2}\right)  
+ u^2(t^2+s^2) \log\left(\frac{-u}{\mu_R^2}\right) \right]
\eea
The values of $g_6,~g_6',~g_8(\mu_R),~g_8'(\mu_R)$ are not predictions of the effective field theory and must be determined by either measurement or by matching to the full theory. We will  explicitly perform the matching below, using the amplitude of the full theory. 

The ``physical'' beta functions of the various couplings can be read off from the amplitude. These are
\bea 
\beta_g &=& 0  \nonumber \\ 
\beta_{g_6} &=& 0  \nonumber \\ 
\beta_{g_6'} &=& 0  \nonumber \\
\beta_{g_8} &=& \frac{41 g^2}{480\pi^2}   \nonumber \\
\beta_{g_8'} &=& \frac{g^2}{240 \pi^2}
\eea
These beta functions are predictions of the effective field theory.

The expected maximum limit of the effective field theory treatment of this matrix element occurs when
\beq\label{limit}
\frac{g E^4}{M^4}\sim 1
\eeq
where $E^4$ here represents any of the kinematic invariants $E^4 \sim s^2, ~t^2, ~u^2$. At these energies the interaction strength becomes large and the EFT treatment fails.
All of the terms in the derivative expansion become relevant, with unknown coefficients. The actual limit of the EFT will either be when new degrees of freedom become dynamically active or at the energy implied by Eq. (\ref{limit}), which ever is lower. 

The key elements of this section are that in the effective field theory treatment: 1) The original coupling $g$ is not renormalized and does not run in the physical sense
and 2) We need to renormalize the couplings of the eight derivative Lagrangian, and these couplings are running couplings.

\section{3. Two-point Function}
In this section, we continue to use the notation of Eq. (\ref{notation}). The Feynman rules are given by the propagator
\be
-\frac{i}{p^2+\frac{1}{m^2}p^4}
\ee
and the four point vertex
\be
-\frac{2ig^2}{M^4}\left[(p_1\cdot p_2)(p_3\cdot p_4)+(p_1\cdot p_3)(p_4\cdot p_2)+(p_1\cdot p_4)(p_3\cdot p_2)\right]
\ee

At one loop, the quantum corrections to the two point function are given by the (tadpole) integral
\be
-\frac{g}{Z_1 M^4}\mu^{4-d}\int \frac{d^dq}{(2\pi)^d}\frac{1}{q^2+\frac{1}{m^2} q^4}\left[p^2 q^2+2(p\cdot q)^2\right]
\ee
In the absence of the four-derivative kinetic term
(i.e. for $m\to\infty$) this is quartically divergent
and is zero in dimreg.

In the general case, setting $d=4-2\epsilon$, it is equal to
\be
i \frac{3}{2}\frac{g}{Z_1}\left(\frac{m}{M}\right)^4 p^2\frac{1}{(4\pi)^2}\left(\frac{1}{\epsilon}+\log{4\pi}-\gamma-\log{\frac{m^2}{\mu^2}}+\frac{7}{6}+O(\epsilon)\right)
\label{2point}
\ee
At one loop, only $Z_1$ receives quantum corrections, since there are no terms proportional to $p^4$. 
However, the $\mu$-dependence in (\ref{2point})
does not correspond to a logarithmic $p$-dependence of the 
2-point function.
This means the $\mu$ dependence of the field renormalization can be reabsorbed once for all without producing any large logs with the physical energy scale of the scattering process.
For example setting $\mu=m$ the logarithm disappears
altogether.

\section{4. The Scattering Amplitude}

Let us now compute the corrections to the four point function.
Since we use signature $-+++$ the Mandel'stam variables are defined as
\bea
s=-(p_1+p_2)^2\\
t=-(p_1+p_3)^2\\
u=-(p_1+p_4)^2
\eea
where all momenta are incoming. For this section we use the notation $Z_2 =1/m^2 $ and work in units with $M=1$, as this matches the previous work using the FRG.\\
In this case one has to consider three Feynman diagrams, that correspond to
the $s$, $t$ and $u$, channels. These are related by crossing symmetry.
In the $s$ channel, the integral one has to evaluate is
\be
\frac{2g^2\mu^{4-d}}{Z_1^2M^8}
\int\!\frac{d^dq}{(2\pi)^d}\frac{N}{(q^2+\frac{1}{m^2}q^4)((q+p)^2+\frac{1}{m^2}(q+p)^4)}
\ee
where $p=p_1+p_2$ and the numerator is
\bea
N&=&\left[(p_1\cdot p_2)(q\cdot(q+p))+(p_1\cdot q)(p_2\cdot(q+p))+(q\cdot p_2)(p_1\cdot(q+p))\right]
\nonumber\\
&&
\times\left[(p_3\cdot p_4)(q\cdot(q+p))+(p_3\cdot q)(p_4\cdot(q+p))+(q\cdot p_4)(p_3\cdot(q+p))\right]\,,
\eea
The other channels only differ by permutations of the external momenta.

Using (\ref{parfrac}), 
the fourth order propagators in the integral can be decomposed in a massless second order propagator and a massive ghost propagator.
This is equivalent to replacing the quartic propagators in the diagrams 
either with the massless or the massive ones,
and summing over all the possible combinations.
In this way, for each channel the correction to the scattering amplitude becomes
$$
\delta\cM=\cM_1-\cM_2-\cM_3+\cM_4\ ,
$$
where $\cM_1$ contains only the contributions of the massless particles,
$\cM_4$ that of the massive ghosts and the other two mixed contributions
with one massive and one massless propagator.
In each partial amplitude we introduce a Feynman parameter,
such that the denominators become (for the $s$ channel)
\bea
\frac{1}{q^2(q+p)^2}&=&\int_0^1 dx\frac{1}{\left(q^{\prime 2}+\Delta_1\right)^2}
\quad\mathrm{with}\quad \Delta_1=x(1-x)p^2\ ;
\\
\frac{1}{q^2[(q+p)^2+m^2]}&=&\int_0^1 dx\frac{1}{\left(q^{\prime 2}+\Delta_2\right)^2}
\quad\mathrm{with}\quad \Delta_2=x(1-x)p^2+xm^2\ ;
\\
\frac{1}{(q^2+m^2)(q+p)^2}&=&\int_0^1 dx\frac{1}{\left(q^{\prime 2}+\Delta_3\right)^2}
\quad\mathrm{with}\quad \Delta_3=(1-x)(xp^2+m^2)\ ;
\\
\frac{1}{(q^2+m^2)[(q+p)^2+m^2]}&=&\int_0^1 dx\frac{1}{\left(q^{\prime 2}+\Delta_4\right)^2}
\quad\mathrm{with}\quad \Delta_4=x(1-x)p^2+m^2
\eea
and $q'=q+x p$.
After some manipulations the numerators become
$$
N=N_0+N_1 (q')^2+N_2(q')^4
$$
where
\bea
N_0&=&x^2(1-x)^2 s^4
\nonumber\\
N_1&=&-\frac{1}{d}\left[(6+d)(x^2-x)+1\right]s^3
\nonumber\\
N_2&=&\frac{1}{4d(d+2)}\left[(d^2+6d+12)s^2+4(t^2+u^2)\right]
\eea
Thus the partial corrections to the amplitude  become, in $d$ dimensions
\bea
\cM_\ell&=&2g^2\frac{1}{Z_1^2}\int \frac{d^d q'}{(2\pi)^d}
\int_0^1 dx \frac{N}{((q')^2+\Delta_\ell)^2}
\\
&=&
\frac{1}{(4\pi)^{d/2}}\frac{2g^2}{Z_1^2}
\int_0^1dx
\left[
\Gamma\left(2-\tfrac{d}{2}\right)\Delta_\ell^{(d-4)/2}N_0
+\tfrac{d}{2}\Gamma\left(1-\tfrac{d}{2}\right)\Delta_\ell^{(d-2)/2}N_1
+\tfrac{d(d+2)}{4}\Gamma\left(-\tfrac{d}{2}\right)\Delta_\ell^{d/2}N_2
\right]
\nonumber
\eea
Finally performing the $x$-integration we obtain for the $s$ channel,
without making any assumptions on the relative size of $s$, $m$, $M$,
\bea
&&\frac{ g^2 m^4 \left(13 s^2+t^2+u^2\right)}{192 \pi M^8
    \epsilon }-\frac{g^2}{5760 \pi^2 s^3 M^8
    } \Bigg\{-3 s^5  \left(41 s^2+t^2+u^2\right)\log
   \left(-\frac{m^2}{s}\right)\nonumber \\
   &&-6 m^4(-s +m^2)^3 \left[ \left(
   s^2+t^2+u^2\right)-2 \frac{s}{m^2}
   \left(-9s^2+t^2+u^2\right)+\frac{s^2}{m^4} \left(41
   s^2+t^2+u^2\right)\right]\log\left(\frac{m^2}{m^2-s}\right)\nonumber \\
   &&
   +s^2 
   m^6 \Big[-2\frac{s}{m^2} \left(352s^2+37(t^2+u^2)-15 \gamma_E\left(13s^2+t^2+u^2\right)\right)\nonumber \\
   &&
   -3 \left(-31s^2+9\left(t^2+u^2\right)\right)+6\frac{m^2}{s}
   \left(s^2+t^2+u^2\right)\Big]
\nonumber \\
   &&
   +6s^{5/2} m^4 \sqrt{4m^2 -s}
    \left(16(6s^2+t^2+u^2)-8\frac{s}{m^2}(16s^2+t^2+u^2)+\frac{s^2}{m^4}(41s^2+t^2+u^2)\right) 
\arccot{\sqrt{\frac{4m^2}{s}-1}}
\nonumber \\
   &&-30 s^3 m^4
   \left(13 s^2+t^2+u^2\right) \log \left(\frac{4\pi\mu^2}{m^2}\right)\Bigg\}\label{channel}
\eea

In this expression we can find a diverging contribution to the $(\partial\phi)^4$ operator, but again the scale parameter $\mu$ appears only in logs divided by the ghost mass $m$, hence the most convenient choice is to set $\mu=m$ for each value of the kinematic variables $s$, $t$ and $u$. The scattering amplitude gains an imaginary part both from the on-shell loops of the massless modes thanks to $\log\left(-\frac{m^2}{s}\right)$ and from the on-shell ghosts in loops when $s>m^2$ in $\log\left(\frac{m^2}{m^2-s}\right)$.\\
The total quantum correction to the four point amplitude is

\bea
&&\frac{5 g^2 m^4 \left(s^2+t^2+u^2\right)}{64 \pi ^2 M^8 \epsilon }
+\frac{g^2}{5760 \pi ^2 M^8}\Bigg\{
\nonumber\\
&&
+\frac{m^4 }{s^2}\Big[-6 m^4 \left( s^2+t^2+u^2\right)
+3 s m^2 \left(-31s^2+9 \left(t^2+u^2\right)\right)
\nonumber\\
&&\qquad\qquad
+2 s^2  \left((352-195 \gamma_E ) s^2-(15 \gamma_E -37)
   \left(t^2+u^2\right)\right)\Big]
\nonumber \\
&&
+6s^{-1/2} m^4 \sqrt{4m^2-s}\big[16m^4(6s^2+t^2+u^2)
\nonumber 
\\
&&\qquad\qquad
-8sm^2(16s^2+t^2+u^2)+s^2(41s^2+t^2+u^2)\big]
\arccot{\sqrt{\frac{4 m^2}{s}-1}}
\nonumber \\
 &&
+\frac{m^4}{t^2} \Big[-6 m^4 \left(s^2+t^2+u^2\right)+3 t m^2\left(-31t^2+9(s^2+u^2)\right)\nonumber \\
   &&
\qquad\qquad
+2 t^2 \left((352 -195 \gamma_E) t^2-(15 \gamma_E -37)\left(s^2+u^2\right)\right)\Big]
\nonumber \\
&&
+6t^{-1/2} m^4 \sqrt{4 m^2-t }\big[16m^4(s^2+6t^2+u^2)
\nonumber 
\\
&&\qquad\qquad
-8t m^2(s^2+16t^2+u^2)+t^2(s^2+41t^2+u^2)\big] \arccot{\sqrt{\frac{4 m^2}{t}-1}}
\nonumber\\
&&
+\frac{m^4}{u^2} \Big[-6 m^4
\left(s^2+t^2+ u^2\right)+3 m^2 \left(-31 u^2+9 \left(s^2+t^2\right)\right)\nonumber \\
&&\qquad\qquad
+2 u^2 \left((352-195 \gamma_E )
   u^2-(15 \gamma_E -37) \left(s^2+t^2\right)\right)\Big]
\nonumber \\
&&
+6u^{-1/2} m^4 \sqrt{4m^2-u}\big[16m^4(s^2+t^2+6u^2)
\nonumber\\
&&\qquad\qquad
-8u m^2(s^2+t^2+16u^2)+u^2(s^2+t^2+41u^2)\big]
\arccot{\sqrt{\frac{4 m^2}{u}-1}}
\nonumber\\
&&
+3 s^2 \left(41 s^2+t^2+u^2\right) \log \left(-\frac{m^2}{s }\right)
\nonumber \\
&&
+3 t^2 \left(s^2+41 t^2+u^2\right) \log \left(-\frac{m^2}{t }\right)
\nonumber\\
&&
+3 u^2 \left(s^2+t^2+41 u^2\right) \log\left(-\frac{m^2}{u }\right)
\nonumber \\
&&
+\frac{6 (u-m^2)^3}{u^3} \log \left(\frac{m^2}{m^2-u}\right)
\left[m^4
   \left( s^2+t^2+u^2\right)-2 u m^2 \left(s^2+t^2-9u^2\right)+u^2  \left(
   s^2+t^2+41u^2\right)\right]
   \nonumber \\
   &&+\frac{6 (t-m^2)^3}{t^3} \log \left(\frac{m^2}{m^2-t}\right) \left[m^4
   \left(s^2+t^2+u^2\right)-2 t m^2 \left(s^2-9t^2+u^2\right)+t^2  \left(s^2+41
   t^2+u^2\right)\right]
   \nonumber \\
   &&+\frac{6 (s -m^2)^3}{s^3} \log \left(\frac{m^2}{m^2-s}\right) \left[m^4
   \left(s^2+t^2+u^2\right)-2 s m^2 \left(-9s^2+t^2+u^2\right)+s^2 \left(41
   s^2+t^2+u^2\right)\right]
   \nonumber\\
   &&+450 m^4
   \left( s^2+t^2+u^2\right) \log \left(\frac{4\pi\mu^2}{m^2}\right)\Bigg\}
\label{4point}
\eea
The arccots can be rewritten as logs, using
\be
\arccot{\sqrt{x-1}}=
\frac{i}{2}\left(\log \left(1-\frac{i}{\sqrt{x-1}}\right)
-\log \left(1+\frac{i}{\sqrt{x-1}}\right)\right)
\ee

\subsection{The $Z_1=0$ case}

It will be instructive to consider the case when there is no two-derivative kinetic term.
Clearly in this case we cannot assume the canonical normalization $Z_1=1$.
In fact, we want to consider the limit when $Z_1$, $m$ and $M$ all go to zero at the same rate.
Defining the dimensionless field $\varphi$ and the coupling $\gamma$ by
$$
\frac{Z_1}{m^2}\phi^2=\varphi^2\ ,\qquad
\frac{Z_1^2}{M^4}\phi^4=\gamma\varphi^4\ ,\qquad
$$
the action (\ref{notation}) becomes 
\beq
\label{notation2}
\cL = \frac{1}{2} m^2\partial_\mu \varphi \partial^\mu\varphi 
- \frac{1}{2}\Box\varphi \Box \varphi 
- \frac{\gamma}{4}( \partial_\mu \varphi \partial^\mu\varphi )( \partial_\nu  \varphi \partial^\nu \varphi)
\eeq
where the field is now canonically normalized with respect to the
four-derivative kinetic term.
Now we can simply set $m=0$.

The calculation of the amplitude follows the steps of the general case
but is much simpler.
The $s$ channel brings the following quantum correction:
\be
\frac{\gamma^2 \left(13 s^2+t^2+u^2\right)}{192 \pi ^2 \epsilon }
+\frac{\gamma^2 \left(3\left(13s^2+t^2+u^2\right) \left( \log \left(\frac{4\pi\mu^2}{-s}\right)- \gamma_E \right)+32 s^2+5\left(t^2+u^2\right) \right)}{576 \pi ^2}+O\left(\epsilon
   ^1\right).
\ee

Defining the renormalized coupling at the scale $s=t=u=\mu_R^2$
by the formula
\be
\gamma(\mu_R)=\gamma-\frac{\gamma^2}{16\pi^2}
\left[
\frac52\left(\frac{1}{\epsilon}+\log\left(\frac{4\pi\mu^2}{\mu_R^2}\right)
-\gamma_E\right)+\frac73
\right]\ ,
\ee
(where the couplings in the r.h.s. are the bare ones)
and exploiting crossing symmetry, we obtain the complete 4-point amplitude
\bea
&&
-\frac{ \gamma(\mu_R)}{2} \left(s^2+t^2+u^2\right)
\nonumber\\
&&
+\frac{\gamma^2}{192 \pi ^2} \bigg[
\log \left(\frac{\mu_R^2}{-s}\right) \left(13 s^2+t^2+u^2\right)
+\log \left(\frac{\mu_R^2}{-t}\right) \left(s^2+13 t^2+u^2\right)+\log \left(\frac{\mu_R^2}{-u}\right) \left(s^2+t^2+13
   u^2\right)\bigg]+O\left(\epsilon ^1\right)
\label{noZ1}
\eea
This agrees with \cite{Tseytlin:2022flu}.

In this case the $\mu_R$-dependence is always associated to the dependence
on the kinematic variables $s$, $t$, $u$.
The physical beta function is
\be
\mu_R\frac{\partial\gamma}{\partial\mu_R}
=\frac{5\gamma^2}{16\pi^2}\ .
\label{betagamma}
\ee

\section{5. Understanding the General Amplitude}

We will study the scattering amplitude in this theory and identify the physical running (or lack of running) of the parameters. The results differ from those given by usual methods. The amplitude calculation is also an instructive example of effective field theory when treated at low energy. Finally we identify a novel (as far as we know) phenomenon of the disappearance of certain operators as one increases the energy.

  Here we will discuss the general case where we start with $Z_1$ and $Z_2$ in principle different from zero. For this section, we will revert to the notation of Eq. \ref{notation}, where $Z_2=1/m^2$ and $g$ is rescaled by a factor of $M^4$.

At one loop there is no renormalization of $Z_2=1/m^2$, as the one loop tadpole diagram only has two factors of the external momentum. We have seen in (\ref{2point}) that the one-loop contribution to $Z_1$ is independent of the momentum. Therefore we can renormalize to $Z_1=1$, and this result will be valid for all energies. From this we see that $Z_1$ is not a running parameter in the amplitude analysis and therefore
\beq
\beta_{Z_1}=0
\eeq 
for all energies. For the rest of this section we set $Z_1=1$.

\subsection{Low energy}

The full result simplifies in the low energy limit. The logarithms involving mass factors can be Taylor expanded in the momentum, so that the only logarithms remaining are of the form $\log -s, ~\log-t, ~\log -u$.  

For $Z_2s\ll Z_1$, we find that the quantum correction is given by
\bea
&&\frac{5  g^2 m^4 \left(s^2+t^2+u^2\right)}{64 \pi ^2 M^8 \epsilon }-\frac{ g^2}{11520 \pi ^2
   M^8} \Bigg\{-900 m^4 \left(s^2+t^2+u^2\right) \log \left(\frac{4 \pi \mu^2}{m^2}\right)
   +30 (30 \gamma_E -11)
   m^4 \left(s^2+t^2+u^2\right)
\nonumber\\
   &&
+6 \Bigg[s^2 \left(41 s^2+t^2+u^2\right) \log \left(\frac{-s}{m^2}\right)
+t^2
\left(s^2+41 t^2+u^2\right) \log \left(\frac{-t}{m^2}\right)+u^2 \left(s^2+t^2+41 u^2\right) \log \left(\frac{-u}{m^2}\right)\Bigg]
\nonumber\\
   &&
   -3  \left(79(s^4+t^4+u^4)+6(s^2t^2+t^2 u^2+u^2s^2)\right)
{-760m^2(s^3+t^3+u^3)}
\Bigg\} \label{smallS}
\eea 

One can see that the logarithm which is proportional to the original interaction, i.e. $s^2+t^2+u^2$, involves $\log ( \mu^2/m^2)$ and is independent of the kinematic variables. This means that we can define a renormalized value of the coupling $g$ by collecting all of the factors which multiply the invariant $s^2+t^2+u^2$ and identifying it with the coupling measured at low energy using the fundamental interaction. 
Then, we find
\beq\label{renormg}
g(\mu) 
=g_B - \frac{5g^2 m^4}{32\pi^2 M^4} \left[ {\frac1{\epsilon} -\gamma_E -\log\left( \frac{4\pi \mu^2}{m^2}\right) +\frac{11}{30}}\right]  \ \ .
\eeq 
Here $g_B$ is the original unrenormalized coupling. 
Now, if we define the beta function by the usual recipe of deriving
with respect to $\mu$ we find
 \beq
\beta_{{g}}^\mu = \mu\frac{\partial g(\mu)}{\partial\mu}
=\frac{5 g^2 m^4}{16\pi^2 M^4}\ .
\label{betamug}
\eeq
However, $g$ does not depend on the energy so the physical beta function is
\beq
\beta_g =0
\eeq
in the LE region. 

The remainder of the amplitude involves powers of energy at order  $E^6 \sim s^3, ~s^2t,...$ and at order $E^8\sim s^4, ~s^2 t^2$. Those of order $E^6$ does not involve any logarithms, while there are logarithms at order $E^4$. A bit of inspection shows that the amplitude is exactly that of the effective field theory given in Eq. \ref{EFTamp}, with the identification
\bea\label{fullEFT}
g &=&g   \nonumber \\
g_6&=& -\frac{53g^2m^2}{384\pi^2 M^2}   \nonumber \\
g_6'&=& -\frac{7g^2m^2}{516\pi^2 M^2}  \nonumber \\
g_8(\mu_R) &=&\frac{79 g^2}{1920\pi^2}   + \frac{41 g^2}{960\pi^2 }\log \frac{\mu_R^2}{m^2}   \nonumber \\
g_8'(\mu_R) &=& \frac{3g^2}{320\pi^2}  + \frac{g^2}{480\pi^2}\log \frac{\mu_R^2}{m^2}  \ \ .
\eea
Whereas in the effective field theory by itself these parameters were unknown, here we see that they are predicted by the full theory. This procedure is referred to as {\em matching} the EFT to the full theory. 

We see that in this region the heavy ghost is not dynamically active and the one loop calculation amounts to integrating it out of the full theory to one loop order. The result is described by an effective field theory, with specific values of the coupling. This is an instructive example of effective field theory reasoning.

\subsection{Above the mass threshold}

Assume that all of the kinematic invariants to be greater that $m^2$ in magnitude, i.e. $(s,~|t|,~|u|)>>m^2$. 
If we use the definition (\ref{renormg}) of the renormalized
coupling defined below the mass threshold,
the amplitude is finite and can be written in the form
\bea
{\cal M} &=&-\frac{g}{2M^4} \left[1{-\frac{17gm^4}{192\pi^2M^4}}\right](s^2+t^2+u^2) \nonumber \\
&& -\frac{g^2m^4}{192\pi^2 M^8}\left[ \log \left(\frac{-s}{m^2}\right)(13s^2 +t^2 +u^2)+  \log \left(\frac{-t}{m^2}\right)(s^2 + 13t^2 +u^2)+ \log \left(\frac{-u}{m^2}\right)(s^2 +t^2 +13u^2)  \right] \ \ .\label{bigS}
\eea

We can instead define the coupling at the (off-shell) renormalization point $s=t=u =\mu_R^2$ by making the finite renormalization
 \beq\label{runningg}
 \bar{g}(\mu_R) =  g +\frac{5g^2m^4}{32\pi^2 M^4}  \left[  \log \left(\frac{\mu_R^2}{m^2} \right) {-\frac{17}{30}}\right]
 \eeq
in which case the amplitude becomes
\bea
{\cal M} &=&-\frac{\bar{g}(\mu_R)}{2M^4} (s^2+t^2+u^2) \nonumber \\
&& -\frac{\bar g^2m^4}{192\pi^2 M^8}\left[ \log \left(\frac{-s}{\mu_R^2}\right)(13s^2 +t^2 +u^2)+  \log \left(\frac{-t}{\mu_R^2}\right)(s^2 + 13t^2 +u^2)+ \log \left(\frac{-u}{\mu_R^2}\right)(s^2 +t^2 +13u^2)  \right] \,.
\label{heamp}
\eea
which agrees with the one calculated in the limit $Z_1=0$,
eq. (\ref{noZ1}).
This is understandable because at high energy the quartic terms in the propagator will dominate of the quadratic terms, and simply ignoring the quadratic terms yields the correct result.

There are a couple of striking observations which can be made from this result. The first is that all of the terms of order $E^8$ and $E^6$ have disappeared from the result. Because the general amplitude of Eq. (\ref{4point}) has many such terms, this requires special cancelations which we will discuss below. The second is that here we {\em can} define a running coupling which removes the potentially large logarithms of the form $\log s/m^2$. We consider the (off-shell) renormalization point $ s=t=u =\mu_R^2$.

For $(s,~|t|,~|u|)>>m^2$ this captures an important part of the quantum correction. There are still logarithms left over, but they are not large. This corresponds to a beta function
\beq
\beta_{\bar{g}} =  \frac{5\bar{g}^2 m^4}{16\pi^2 M^4}
\label{betabg}
\eeq
The $17/30$ in the formula for Eq. \ref{runningg} amounts to an optional threshold correction matching the amplitude above and below the threshold. 

The disappearance of the $E^8$ and $E^6$ terms appears initially surprising. There are many such terms with factors such as $s^4,~ t^4,...$ and $s^3,~t^3,...$ in the general result. However, we can begin to see that there are cancelations by looking at the logarithmic type terms which arise at the highest order, $E^8$. We recall the result of Passarino and Veltman that all one loop diagrams can be expressed in terms of factors of the scalar tadpole, bubble, triangle and box diagrams. Here only the tadpoles and bubbles contribute. The tadpoles do not depend on the external momenta and do not give kinematic logs. These kinematic factors inside the logarithms come from the scalar bubble diagrams, which have the form
\beq\label{bubble}
I_2(m_1,~m_2,~q^2) = \frac1{16\pi^2}\left[  \frac1{\epsilon} + \gamma -\log 4\pi - \int_0^1 dx \log \left(  \frac{xm_1^2+(1-x)m_2^2 -q^2x(1-x)}{\mu^2}\right)    \right]   \ \ .
\eeq
The logarithmic integral has the form
\bea
 \int_0^1 dx \log \left(  \frac{xm_1^2+(1-x)m_2^2 -q^2x(1-x)}{\mu^2}\right)  &=& \log \left(-\frac{q^2}{\mu^2} \right)-2~,~~~~~~~~~~~~~~~~~~~~~~~~~~~~~~~~~~~~~~~~~~~~~m_1=m_2 = 0 \nonumber \\
 &=& \log \frac{m^2}{\mu^2} + \left( 1-\frac{m^2}{q^2} \right) \log \left( 1-\frac{q^2}{m^2} \right)-2~,~~~~~~~~~~~~~~~ m_1=0, ~m_2=m  \nonumber \\
 &=& \log \frac{m^2}{\mu^2}  + \sqrt{1-\frac{4m^2}{q^2}} \log \left( \frac{\sqrt{1-4m^2/q^2}+1}{\sqrt{1-4m^2/q^2}-1} \right)   -2~,~~m_1=m_2=m   \ \ .
 \nonumber\\
 \eea
The reader can see these logarithmic factors in the general amplitude. Moreover, one can see that there are common factors preceding these logs and the result involves the combination
\beq\label{cancelation}
I_2(0,~0,~ q^2) -2I_2(0,~m, ~q^2)+ I_2(m,~m,~q^2)      \ \ .
\eeq
The divergences and the factor of $\log \mu^2$ cancel with this combination leaving a finite result as observed. In the low energy region, the latter two components of this expression go to constants, and only $I_2(0,0,q^2) $ gives kinematic logarithms. This leads to the log dependence found in the LE/EFT limit in Eq. \ref{EFTamp}. However at high energy each of the components involves equal factors of $\log(-q^2)$, and the leading energy dependence of this combination will cancel. This leads to the vanishing of the terms of order $E^8$ at high energy. It requires detailed work to verify that the remaining terms of order $E^6$ also cancel, but the general idea is the same. The amplitude that starts out containing orders $E^4,~E^6,~E^8$ at low energy ends up at order $E^4$ only at high energy. To the best of our knowledge, this vanishing of high order energy depencence which were required within the effective field theory is a novel effect not previously described in the literature.

\subsection{The very high energy region} 

The results of the preceding section hold for all energy scales such that the mass can be ignored.
However, in the Introduction we distinguished an intermediate from a high energy scale.
In the high energy region a puzzling situation now presents itself.
From equations (\ref{betagamma}) or (\ref{betabg})
one sees that for $g<0$ the coupling is asymptotically free
(in agreement with earlier calculation \cite{Tseytlin:2022flu}).
However, it appears that a focus on the coupling constant is insufficient. Even if the coupling constant is running logarithmically to an asymptotically free fixed point, the amplitude itself is blowing up with energy. 

At high enough energy, the one-loop scattering amplitude will become greater than unity. This occurs when the kinematic invariants $s$, $t$, $u\sim E^2$ are of order
\beq
\frac{ g(E) E^4}{M^4} \sim 1  \ \ .
\eeq
A logarithmic decrease in the coupling does not offset the power-law growth. 
This behavior puts the notion of asymptotic freedom in question. We leave a more detailed discussion to
a separate publication,

 \section{6. Different definitions of running parameters}
Here we return to our introductory point that there are different flavors of renormalization group techniques. In turn, we  address the three which we highlighted.

\subsection{Physical beta functions}

 Part of the renormalization program is the measurement process. Because we do not know the bare couplings, we need to measure the parameters. 
 Within a given scheme, this involves 
 measurements at a renormalization scale $\mu_R$. When the physical amplitude depends on
 the energy in a particular way, then measurements at different renormalization scales will lead to different values of the coupling. What we refer to as the physical beta function describes how the coupling changes with $\mu_R$. 
 
 The important point here is that this procedure studies the physical amplitudes and their
 dependence on the energy. We have used this method in the preceding analysis. In our work, the renormalization 
 scheme involved identifies the couplings at the symmetric point $s=t=u=\mu_R^2$.

\subsection{Using divergences or $\log \mu$ to define running couplings}

 Rather than calculate physical amplitudes, we often just look at the renormalization constants required for renormalizing the parameters. In dimensional regularization, these depend on $\log \mu^2$ where $\mu^{(4-d)}$ is the parameter introduced to keep the dimensions of Feynman integrals a constant. The dependence on $\log \mu$ comes along with the $1/\epsilon$ of the renormalization constant and hence $\log \mu^2$ appears in the renormalization in the same way every time the coupling appears. Then beta functions can be calculated from  $\mu \frac{\partial}{\partial \mu}$. In regularization schemes with a cutoff, beta functions can be found using  $\Lambda \frac{\partial}{\partial \Lambda}$.

In mass-independent renormalization schemes, this procedure will identify the physical running. Because there are no other dimensionful factors around, the factors of $\log \mu$ will always be accompanied by factors of $\log E^2$, as in $\log s/\mu^2, ~\log t/\mu^2...$  when applied in an amplitude.

However, there are a couple of ways that this could go wrong when there are factors of masses around. The logarithm could involve $\log m^2/\mu^2$ or $\log m^2/\Lambda^2$. In this case, the $\log \mu^2$ or $\log \Lambda^2$ dependence does not correlate with any energy dependence in the physical amplitude. It is just a constant and disappears when the renormalized coupling is defined. 

A somewhat more unusual case where this method fails is in running parameters which are not associated with $1/\epsilon, ~\log \Lambda, ~\log \mu$. An example of this is seen in our scattering amplitude. The low energy analysis using the EFT shows that the couplings at order $E^8$ are running couplings at one loop, see Eq. \ref{EFTamp}. This occurs for the usual reason, with the EFT involving mass-independent renormalization, and the quantum corrections at order $E^8$ involving $1/\epsilon - \log E^2/\mu^2$. However in the full theory, these terms do not involve any renormalization, nor factors of $\log \mu$, as can be seen in Eq. \ref{fullEFT}, as they are finite predictions. Yet since the amplitudes match the EFT analysis exactly, $g_8, ~g_8'$ are physically running parameters. 

\subsection{One loop beta functions from the FRG}

In \cite{Buccio:2022egr} the Lagrangian was parametrized as in (\ref{action1})
and the running of $Z_1$, $Z_2$ and $g$ was calculated using the full FRG.
The coupling then depend on a scale $k$ that has the meaning of an IR cutoff.
This calculation goes beyond the one loop approximation,
because the couplings in the r.h.s. of the FRG equation are treated
as running couplings. This kind of ``RG improvement'' 
amounts to a resummation of infinitely many diagrams.
In order to compare with the amplitude calculation,
we have to downgrade those results to the one loop approximation.
This is easily achieved by neglecting the RG improvement.

Assuming that the field has dimension of mass,
one arrives at the following beta functions
\bea\label{FRGresults}
k\partial_k Z_1&=&
-\frac{Z_1+2 k^2 Z_2}{16\pi^2(Z_1+k^2 Z_2)^2}\,g k^4
\label{bZ1}
\\
k\partial_k Z_2&=&0
\\
k\partial_k g&=&\frac{5(Z_1+2k^2 Z_2)}{32\pi^2(Z_1+k^2 Z_2)^3}\,g^2 k^4
\label{bg}
\eea
With dimensionless field the beta functions are the same,
but of course the dimension of the couplings is different
and so are the powers of $k$ in the r.h.s.
These one loop beta functions differ from the full ones of \cite{Buccio:2022egr},
but the qualitative features of the RG flow remain the same.

In order to study the RG flow one has to make the couplings
dimensionless, multiplying them by powers of $k$.
One then obtains different results depending on the dimension of the field.
When the field has dimension one, one finds the low-energy gaussian fixed point
corresponding to the free theory with two-derivative propagator,
as well as a bunch of other nontrivial fixed points.
When the field is dimensionless, one finds the high-energy gaussian fixed point
corresponding to the free theory with four-derivative propagator,
as well as a bunch of other nontrivial fixed points.
The two descriptions can be seen as two charts on a manifold,
related by a well-defined coordinate transformation.
In each chart, one of the free fixed points sits in the origin,
while the other lies asymptotically at infinity.

The nice feature that was observed in \cite{Buccio:2022egr}
is that the running of $Z_1$ can be absorbed in a redefinition of the field,
and then the dimension of the field changes continuously
from zero near the UV fixed point to one near the IR fixed point.

\subsection{Comparisons}

We are now ready to compare  the physical running with
the results of the FRG and the $\mu$-running.

\smallskip

We begin by comparing the physical running of $g$ to the $\mu$-running.
At low energy, below the mass $m$, the amplitude does not give a physical running for $g$:
\beq
\beta_g =0,~~~~~~~~E\ll m \ \ .
\eeq
{This is in disagreement with the $\log\mu$ derivative approach, which predicts a logarithmic running, see eq.(\ref{betamug}).
This $\beta$ function comes out from  the $\log m^2/\mu^2$ in (\ref{renormg}). Here $\mu$ is an unphysical parameter which disappears from all physical reactions after renormalization. The apparent running of the coupling $g$ arises from taking the negative logarithmic derivative of correction (\ref{renormg}) with respect to $\mu$. This often is appropriate in other settings because in mass independent renormalization schemes the logarithmic factor is $\log q^2/\mu^2$ (where $q$ is some kinematic energy factor) so that taking the derivative with respect to $\mu$ reveals the dependence of the amplitude on the kinematic variables $\log q^2$. However here there is no dependence on any kinematic variable. If we perform renormalization at any kinematic scale below the mass threshold, it remains that value as long as the ghosts stay frozen. Of the two definitions of running given in dimensional regularization, the physical one matches the results from the EFT in section 2, where we did not observe any quantum corrections to the coupling $g$. 

A well known example can illustrate this point. In QED, the vacuum polarization correction involving a top quark loop yields the correction 
 \beq\label{QEDlow}
 \Pi(q^2) =\frac { \alpha}{3\pi} \left[ \frac1{\epsilon} -\gamma + \log 4\pi -\log\frac{m_t^2}{\mu^2} +\frac{q^2}{5m_t^2} +...\right]
 \eeq
 at low energy. However, despite the dependence on $\log \mu$, this does not imply that the top quark loop contributes to the running of the electric charge at low energy. The top quark makes a contribution to the running of $\alpha$ only at energies above $m_t$, where the logarithmic factor involves $\log q^2$ instead of $\log m_t^2$.

There is agreement in the running of $g$ in the high energy region for energies above the mass $m$. Here the beta function describing the running coupling in the amplitude is
 \beq
\beta_{{g}} =  \frac{5{g}^2m^4}{16\pi^2 M^4} ,~~~~~~~~{E\gg m}\ \ .
\eeq
This indeed agrees with equation (\ref{betamug}),
and does not change if we add a finite piece as in (\ref{runningg}).

Again, we can see the relevant physics in the simpler case of the QED vacuum polarization. While the low energy form of the function is given in Eq. \ref{QEDlow}, the high energy version of this is
 \beq\label{QEDhigh}
 \Pi(q^2) =\frac { \alpha}{3\pi} \left[ \frac1{\epsilon} -\gamma + \log 4\pi -\log\frac{m_t^2}{\mu^2}  -\log \frac{-q^2}{m_t^2} +...\right]
\eeq
 Asymptotically the mass cancels out, and we could have performed the renormalization using a mass-independent scheme. However, since we previously chose to renormalize the electric charge at low energy, absorbing the $\log m_t^2/\mu^2 $ factor into the coupling, the latter logarithm is potentially a large logarithm and should be resummed in a running coupling constant. The top quark contribution to the electric charge is constant at low energy and runs at high energy. 
 
 This is the same behavior as is revealed in the running of $g$. We can see this relatively simply in the calculation. The loop integral in this model is proportional to 
 \bea
 I_{\mu\nu\alpha\beta} &=&m^4\int \frac{d^dp}{(2\pi)^d} \frac{p_\mu p_\nu ( p-q)_\alpha (p-q)_\beta}{[m^2 p^2 - p^4][(m^2 (p-q)^2 - (p-q)^4]}  \nonumber \\
 &=& F(q^2) (\eta_{\mu\nu}\eta_{\alpha\beta} +\eta_{\mu\alpha}\eta_{\nu\beta} + \eta_{\mu\beta}\eta_{\nu\alpha}) +{\rm order~ q ~terms }\ .
 \eea
 The only divergence appear in the first term $F$. We can simply evaluate this divergence by taking the trace of this integral
 \bea
 \eta^{\mu\nu}\eta^{\alpha\beta} I_{\mu\nu\alpha\beta} &=& m^4 \int\frac{d^dp}{(2\pi)^d} \frac1{[m^2 - p^2][(m^2  - (p-q)^2]}   \nonumber \\
  &=& = m^4 I_2(m,m,q) =  d(d+2) F+....
  \eea
where $I_2$ is given in Eq. \ref{bubble}.  This is just the scalar bubble diagram, which along with the divergence carries the $\log m^2/\mu^2$ factor at low energy and $\log q^2/\mu^2$ at high energy, just as we have seen above. There are other logarithms possible in the amplitude, but this one is tied to the renormalization of the coupling at low energy and the running at high energy.  The FRG and the dimensional regularization of the running agree because it is uniquely the bubble diagram which determines the divergent factor in this calculation.

We now compare the physical running of $g$ to the FRG results.
At low energy  the FRG running is power-law:
\beq
\beta_g = \frac{5(Z_1+2k^2/m^2)}{32\pi^2 (Z_1+k^2/m^2)^3} \frac{g^2k^4}{M^4}  \to\frac{5{g}^2k^4}{32\pi^2 M^4} ,   ~~~~\mathrm{for}~k\ll m\ \ .
\eeq
 This beta function
rapidly runs to zero at lower energies, asymptoting to the constant value of $g$ found in the amplitude calculation. 

On the other hand, at high energy the FRG gives
 \beq\label{FRGg}
 \beta_g = \frac{5(Z_1+2k^2/m^2)}{32\pi^2 (Z_1+k^2/m^2)^3} \frac{g^2k^4}{M^4}  \to\frac{5{g}^2m^4}{16\pi^2 M^4} ,   ~~~~~~~k\gg m\ \ .
\eeq
which agrees both with the physical running and the $\mu$-running.

Between these two limits, the behavior of the amplitude is more complicated, and if one tries to define a running coupling as in (\ref{runningg}), namely isolating the coefficient
of $s^2+t^2+u^2$ in the full amplitude (\ref{4point}) and
setting $s=t=u=\mu_R^2$,
it turns out to be impossible to unambiguously identify it. 
Hence the definition of a physical running is only meaningful in the asymptotic regions, where different power-laws are clearly separated.
The standard, conventional way of joining them is to assume that $g$ does not run
all the way up to the mass $m$ and to match this to the high-energy
logarithmic behavior (\ref{heamp}) via (\ref{runningg}).
This is shown by the black dashed line in Fig.\ref{fig:pot}.

We note that whereas the beta function becomes universal (scheme-independent)
at high energy, the relation between the low-energy value of the coupling and
its high-energy behavior is not.
By choosing a different constant in the bracket in (\ref{runningg}) we can
change the offset between the low- and high-energy parts of the curve in Fig.\ref{fig:pot}
and shift up or down the part of the curve above the threshold $k/m=1$.

The same effect can also be obtained by using
a different renormalization point.
In the UV, the non-polynomial dependence on the kinematical variables of the terms of the amplitude proportional to $s^2$, $t^2$ and $u^2$ is given by $\log(-s/\mu)$, $\log(-t/\mu)$ or $\log(-u/\mu)$. 
Thus, if we choose to renormalize at $s=at=bu=\mu_R^2$ with $a$ and $b$ fixed constants, the coupling gets shifted by $-\log(ab)$, at the price of having $\log \left(\frac{-ta}{\mu_R^2}\right)$ and $\log \left(\frac{-ub}{\mu_R^2}\right)$ in eq. (\ref{heamp}). 
In this work we have used the symmetric point $a=b=1$
and the definition (\ref{runningg}),
because these best capture the behavior of the amplitude and
are best suited for comparison to the FRG,
but we stress that these are arbitrary choices.

\begin{figure}[ht]
\begin{center}
\includegraphics[scale=0.7]{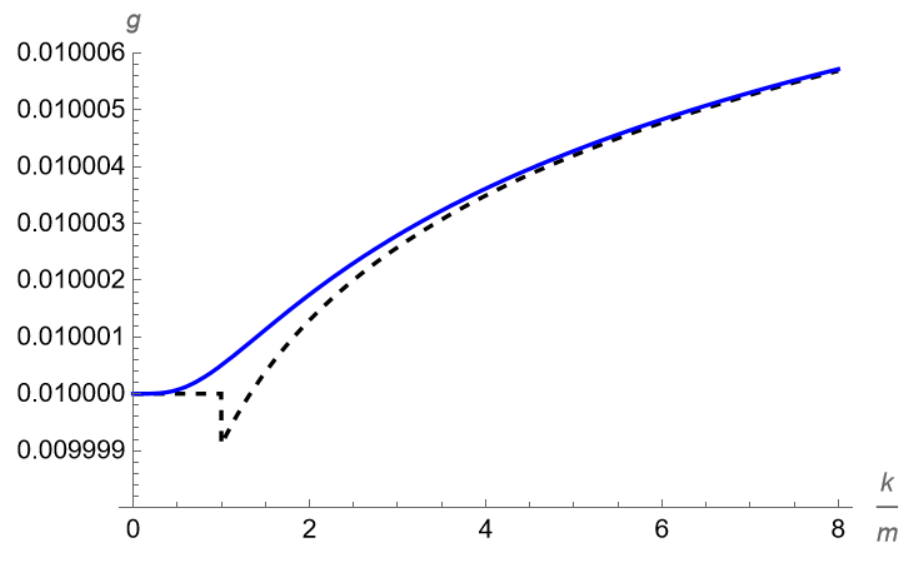}
\caption{The running coupling calculated from the FRG (blue continuous curve)
and the one obtained by matching the low- and high-energy physical running
(black dashed). They have been calculated here for the same low energy limit
$g=0.01$.
}
\label{fig:pot}
\end{center}
\end{figure}

In on-shell configurations, choosing the parameters $a$ and $b$ is equivalent to fixing the scattering angle. This angle should be held fixed along the running from the IR to the UV regime, otherwise one could observe different beta functions and consequently different runnings of the coupling.
If one allows the scattering angle to also depend on $s$, the amplitude is no longer 
described by the universal running coupling, as demonstrated by the example
of peripheral scattering of Appendix A.

The FRG gives a continuous interpolation for the running of the quartic coupling,
and can separately account for the six- and eight-derivative couplings
that will inevitably be generated.

In order to compare the RG trajectory of the coupling in the FRG
with the trajectory of the physical coupling, we have to
make an identification of the argument of the former,
which is an arbitrary cutoff scale $k$,
with the argument of the latter,
which at the symmetric point is $\sqrt s$.
If we just put $k^2=s$, and we adjust the initial conditions so that the
two trajectories have the same IR limit $g(0)$,
then in the UV limit they differ by a small offset.
This can be fixed by choosing $k=\sqrt s/\xi$, where 
$\xi=e^{25/40-17/60}\approx 1.4$.
This is illustrated again in Fig.\ref{fig:pot}.

\medskip

The other running parameter within the FRG is $Z_1$. In the notation of this section, the general expression was
 \beq
 \beta_{Z_1} =-\frac{Z_1+2k^2/m^2}{16\pi^2(Z_1+k^2/m^2)^2} \frac{g k^4}{M^4}    \ \ .
 \eeq
This is in disagreement with the amplitude calculation, for which $Z_1$ does not run
at all energies
\beq\label{ampZ}
\beta_{Z_1}= 0
\eeq

If we had defined the running of $Z_1$ not by the dependence on energy or on renormalization scale, but by the dependence of the counterterm on the unphysical parameter $\mu$ which appears in dimensional regularization, we would have identified
 \beq\label{muZ}
 \beta^\mu_{Z_1} =\frac{ 3}{16\pi^2} \frac{gm^4}{M^4}   \ \  .
 \eeq
 We noted that the asymptotic form at large $k$ of the FRG result was
 \beq
 \beta_{Z_1} = \frac{gm^2k^2}{M^4}  +\frac{ 3}{16\pi^2} \frac{gm^4}{M^4}  +...
 \eeq
 which, if one disregards the power-law running, would agree on the logarithmic running. So in this case {the issue concerning the proper definition of the beta function is not limited to a given kinematical domain, but whether considering $Z_1$ a 
{physically} running coupling at all.
 
  The one-loop correction to the kinetic energy term was found to be
 \beq
 \frac{3gm^4}{16\pi^2M^4}p^2 \left[\frac1{\epsilon} -\gamma +\log 4\pi -\log \frac{m^2}{\mu^2} +\frac{7}{6} \right]  \  \ .
 \eeq
 {The portion to focus on is again the $\log m^2/\mu^2$ and what is going on is very similar to the low energy regime of $g$.} If we perform wavefunction renormalization at any kinematic scale, setting $Z_1=1$, it remains that value at any other scale. Taking the derivative with respect to $\mu$ 
does not give us physical information in this case. 

The other issue is that of power-law corrections found within the FRG.
In the case of $Z_1$ this is a less significant issue than for $g$,
since $Z_1$ is a redundant coupling and is not associated directly with
any scattering process.
Another way to say this is that in a two-point function
the only invariant scale is $p^2$, which on shell is just equal to the pole mass $m^2$.
Nevertheless, we can interpret the difference between the beta functions 
computed here and those coming from the FRG as follows.
In this paper we have perturbed around a generic free theory containing
both kinetic terms, which is not a fixed point in general:
only the theories with $Z_1=0$ or $Z_2=0$ are fixed points.
There is a trivial running with $p$ that goes from the one to the other,
since the quartic term dominates in the UV and the quadratic one
dominates in the IR,
but the dimension of the field remains fixed and does not enter in any of our conclusions.
However, the canonical dimension of the field at a free fixed point is fixed:
it is one at the two-derivative fixed point and zero at the four-derivative one.
In the FRG this is correctly taken into account.
Within the context of the FRG, the power running of $Z_1$ with scale $k$ is necessary to
correctly interpolate between the low-energy and high-energy Gaussian fixed points.

\section{7. Discussion}

We have explored the running of couplings in a simple model with higher derivative interactions and kinetic energy. The scattering amplitude reveals what we are calling the ``physical'' running, as it describes the running parameters seen in physical processes. This differs from some other definitions of running couplings using different methods, and we have used explicit calculations to illustrate these differences.

Some of the lessons from this work can be summarized as follows;
\begin{enumerate}

\item {Physical running couplings can only be defined far from mass thresholds,
and}
there are different patterns of running above and below the threshold. In our case, the coupling $g$ does not run below the threshold and runs logarithmically above it. Effective Field Theory is useful in understanding the low energy region.

\item Power-law running is not seen in the physical amplitudes. Instead, in the EFT regime, the effects which depend on higher powers of the kinematic invariants are organized as higher order operators in an effective Lagrangian.
These higher order operators disappear altogether above the mass threshold (operator ``melting''). 

\item Alternate methods of defining running couplings using $\Lambda \frac{ \partial}{\partial \Lambda}$ , $k \frac{ \partial}{\partial k}$ or $\mu \frac{ \partial}{\partial \mu}$ (where $\Lambda, ~k, ~\mu$ refer to UV cutoffs, IR cutoffs or the dimensional regularization auxiliary scale) sometimes yield running behavior which is not seen in physical processes. In our case, this is found in the coupling $Z_1$. The culprit is factors of $\log m^2/\Lambda^2$ etc, which does not involve any of the kinematic invariants and hence does not change with the energy scale of the physical reaction.

\end{enumerate} 

The last point calls into question the utility of these alternate methods if the results are not reflected in physical amplitudes.
In the case of the $\mu$-running,
terms like $\log(\mu^2/m^2)$ in the amplitude do not correspond to physical running.
Sometimes such terms arise below the mass threshold
and are replaced by genuine running $\log(\mu^2/-q^2)$ above threshold, as we have seen in the example of the 
top quark contribution to the running of the electromagnetic coupling. 
However, there may be other examples of beta functions where
this is not properly accounted for.
This seems to be the case for example in the nonlinear sigma models
\cite{Hasenfratz:1988rf,Percacci:2009fh}
and in Quadratic Gravity \cite{ft1,Avramidi:1985ki}
whose beta functions need to be recalculated. 
We will report these results separately in joint work with G. Menezes.

The FRG deserves a separate discussion.
 One way of viewing it is as a method to compute the effective action,
in alternative to the path integral.
One starts from a given form of the scale-dependent effective action,
presumed valid at some UV scale $\Lambda$, 
then includes the effect of quantum fluctuations
with momenta between $\Lambda$ and a lower scale $k$,
by integrating the FRG from $\Lambda$ to $k$.
When one integrates down to $k=0$, the full effective action is obtained,
and it contains all the information about all scattering amplitudes.
For an explicit example of such a calculation in the context of $\phi^4$ theory we refer to \cite{Codello:2015oqa}.
In this use, $k$ is by itself an unphysical variable. If one were to calculate at some non-zero value of $k=k_*$ one should then add in the quantum corrections that come from the region of $k=0$ up to that value of $k=k_*$, which were excluded by the cutoff. The physical amplitudes would be independent of the choice to work at any non-zero value of $k_*$. 
That is the origin of the renormalization group equations - describing how the coupling must change with the cutoff in order to hold the physical properties fixed.

The second way to use the FRG is based on giving a physical meaning to $k$
(a so-called ``cutoff identification'').
It rests on the ability of the FRG to account for decoupling phenomena,
that is one of its main strengths (see e.g. \cite{Berges:2000ew}).
If some physical variable $F$, with dimension of mass,
is the only mass scale of the system,
and if it enters in loop calculations in the same way as a mass or an IR cutoff,
then the $F$-dependence  of the effective action (at $k=0$)
will be the same as the $k$-dependence of the running effective action.
This is because for $k<F$ the decoupling theorem implies that
the effective action becomes independent of $k$.
We refer to \cite{Reuter:2003ca} for a clear exposition of this logic.

In our calculations, the identification of the cutoff with external momenta
sometimes works and sometimes not.
Consider first the corrections to the two-point function.
If one had calculated the tadpole diagram with cutoff regularization rather than dimensional regularization, the tadpole integral is quadratically divergent and one would find a result that has the form
\beq
\frac{3gm^4}{16\pi^2M^4}p^2 \left[c\Lambda^2 -\log \frac{m^2}{\Lambda^2} +....\right]  \  \ .
\eeq
where the constant $c$ would depend on how the cutoff is implemented. In this case $\Lambda$ is an UV cutoff so that quantum corrections below $\Lambda$ are included. However, in this case we know that, treated as a regularization scheme, both the power-law and logarithmic dependence on $\Lambda$ disappear after renormalization and we again have $Z_1=1$. So in this case, tracing the cutoff dependence of the counterterm
does not say anything about the two-point function.
In the FRG calculation the tadpole gives rise to quadratic running of $Z_1$.
However, the external momentum does not enter in any way in the tadpole integral
and so this not a case where one can trade the external momentum dependence
for $k$-dependence.
 
In the running of the coupling $g$ we see both outcomes. At low energy the power-law running found in the FRG is not observed in the amplitude. 
It is an aspect of the threshold behavior, 
interpolating between a constant in the IR limit and the logarithmic
behavior at the high energy.
The threshold behavior of the couplings in FRG is not universal,
and in any case there is no definition of physical running to compare with
in that regime.

At high energy, the dependence on $\log k^2$ mirrors correctly the dependence of the amplitude on $\log E^2$, and gives the correct beta function. 
This is due to the fact that
for fixed ratios $t/s$ and $u/s$, and in the limit when $m/s\to 0$,
the amplitude depends only on a single mass scale $s$,
which enters in the denominators of the loop intergrals in a way that
is reminiscent of an IR cutoff.
Thus, in this regime, the $k$-dependence of the running coupling
correctly reflects the $s$-dependence of the amplitude.
At energies close to $m$, the amplitude becomes a complicated function
of $s$ and $m$ and no RG calculation exactly reproduce the amplitude.

The model discussed in this paper reiterates
several points made by one of us in the past
\cite{Anber:2010uj,Anber:2011ut,Donoghue:2019clr},
but there are also some new aspects.
The disappearance of the higher order operators of the low energy EFT
is expected when the model is UV completed in a linear $U(1)$ sigma model,
but surprisingly also happens when the four-derivative kinetic term
becomes important.
This offers a glimpse of how, in a derivatively coupled theory, one could transition from the
low energy EFT regime to an asymptotically free (and possibly asymptotically safe) regime.
In principle, this could provide an alternative UV completion
to the $U(1)$ linear sigma model, mentioned in the Introduction.

This kind of behavior may be extended also to gravitational theories.
For example, it raises the possibility that at least some of these higher order operators, such as those of order $R^3$ should not be used above certain thresholds,
because the coefficients of the higher order operators vanish.
Our model seems to enter a strong coupling regime at very high energy.
This is because the powers of momentum of the interaction
overwhelm the logarithmic decrease of the coupling.
We have not discussed the physics of this regime in this paper,
but we plan to return to it in the future.

\section*{Acknowledgments} 
JFD thanks the organizers of the SISSA-IPFU workshop on Quantum Effective Field Theory and Black Holes, which led to his involvement in this project
and acknowledges partial support from the U.S. National Science Foundation under grant NSF-PHY-21-12800. 
JFD also thanks Gabriel Menezes, Cliff Burgess and Bob Holdom
and RP thanks Gian Paolo Vacca for useful conversations.

\break

\section*{Appendix - Peripheral Scattering}

The peripheral scattering limit is that of large $s$ and $t\sim 0$. Although this section will only be peripheral to our main discussion (pun intended), it does have an interesting feature which we will comment on in the discussion.  

In this case, the $s$ and $u$ channels give the similar contributions to the scattering amplitude, since on shell $u=-s$ and the Mandel'stam variables appear only quadratically or in the Log in the dominant terms in the high energy limit. 

\be
\frac{7 g^2 m^4 s^2}{48 M^8 \pi ^2 \epsilon }
+\frac{g^2 m^4}{576 M^8\pi ^2} \left\{(74-84 \gamma_E )s^2
+42s^2\left[\log \left(\frac{4\pi\mu^2}{s}\right)+\log \left(-\frac{4\pi\mu^2}{s}\right)\right]\right\}+O\left(\epsilon ^1\right)
\ee
On the other hand, the $t$ channel is a bit more subtle, since the terms powers of $t$ in the denominator could give some divergences. However, if we expand (\ref{channel}) with $t$ and $s$ exchanged at small $t$, all the divergent terms are actually zero and we obtain
\bea
&&
\frac{ g^2 m^4 s^2 }{96 \pi ^2 M^8\epsilon }+\frac{g^2 m^4 s^2 \left[6  \log \left(\frac{4 \pi  \mu ^2}{m^2}\right)-(6 \gamma -5) \right]}{576 \pi ^2M^8}
\nonumber\\
&&+\frac{ g^2 m^2 s t  }{96 \pi ^2 M^8 \epsilon }+\frac{g^2 m^2 s t  \left[7 s +12 m^2  
   \log \left(\frac{4 \pi  \mu ^2}{m^2}\right)+2 m^2 (-6 \gamma   +5)\right]}{1152 \pi ^2 M^8  }
\nonumber\\
   &&+\frac{7  g^2 t^2 m^4
   }{96 \pi ^2 M^8 \epsilon }+\frac{g^2 t^2 \left[-6  s^2
   \log \left(-\frac{t }{m^2}\right)  +9  s^2
     +35 m^2 s   +420  m^4   \log \left(\frac{4 \pi  \mu
   ^2}{m^2}\right)-m^4(420  \gamma    +140 ) \right]}{5760 \pi ^2 M^8}
   \nonumber\\
   &&+O\left(t^3\right)
\eea
Anyway the first line is clearly dominant.
Hence the one loop quantum corrections in peripheral scattering are
\bea
\frac{5 g^2 m^4 s^2 }{32\pi ^2 \epsilon M^8 }
+\frac{g^2m^4s^2}{576 M^8\pi ^2} \left\{(79-90 \gamma_E )
+42\left[\log \left(\frac{4\pi\mu^2}{s}\right)+\log \left(-\frac{4\pi\mu^2}{s}\right)\right]+6  \log \left(\frac{4 \pi  \mu ^2 }{m^2}\right) \right\}+O\left(t\right)
\eea

After renormalization, the amplitude has the form in the present notation
\beq
{\cal M} = \frac{g s^2}{M^4} \left[  1  + \frac{7 gm^4}{96\pi^2 M^4} \left(  \log \left(\frac{-s}{m^2}\right) +\log \left(\frac{s}{m^2}\right)   \right)  +\frac{79gm^4}{576\pi^2M^4}   \right]  +{\cal O}(st) \ \ .
\eeq
For this process one can defined a running coupling 
\beq
{\tilde g} (\mu_R)= g +  \frac{7 g^2 m^4}{48\pi^2 M^4}\log \frac{ \mu_R^2}{m^2} +\frac{79g^2m^4}{576\pi^2M^4}  \nonumber
\eeq
when renormalizing at the scale $s=\mu_R^2$, where again the $79/576\pi^2$ factor is optional. This removes the potentially large logarithms, and carries the beta function
\beq  
\beta_{\tilde g} = \frac{7g^2 m^4}{24\pi^2 M^4}  \ \ . \nonumber
\eeq
It is interesting that one can define a physical beta function in this region, 
yet it is different from that found when all the kinematic variables are large.

We can understand this in the following way.
The universal beta functions that one calculates from perturbation theory
are only universal as long as one considers processes that depend on a single momentum scale.
This is the case, for example, for $2\to2$ scattering at a fixed angle:
the ratios of the Mandel'stam variables are fixed and the
amplitude depends just on $s$.
In the case of peripheral scattering we are changing the scattering angle
together with the energy, and the amplitude is not a function of $s$ alone.
While there is no guarantee that a running coupling can be defined in this setting, it appears possible in the one loop calculation.

\end{document}